\documentclass[conference]{IEEEtran}
\IEEEoverridecommandlockouts
\usepackage{cite}
\usepackage{amsmath,amssymb,amsfonts}
\usepackage{algorithmic}
\usepackage{graphicx}
\usepackage{textcomp}
\usepackage{xcolor}
\usepackage{listings}
\usepackage{longtable}
\usepackage{booktabs}
\usepackage{url}

\def\BibTeX{{\rm B\kern-.05em{\sc i\kern-.025em b}\kern-.08em
    T\kern-.1667em\lower.7ex\hbox{E}\kern-.125emX}}
\begin{document}

\title{News Sentiment Embeddings for Stock Price Forecasting}

\author{\IEEEauthorblockN{Ayaan Qayyum}
\IEEEauthorblockA{\textit{Department of Electrical \& Computer Engineering} \\
\textit{Rutgers University}\\
New Brunswick, NJ, USA \\
Email: aaq18@rutgers.edu }
}

\maketitle

\begin{abstract}
This paper will discuss how headline data can be used to predict stock prices. The stock price in question is the SPDR S\&P 500 ETF Trust, also known as SPY that tracks the performance of the largest 500 publicly traded corporations in the United States. A key focus is to use news headlines from the Wall Street Journal (WSJ) to predict the movement of stock prices on a daily timescale with OpenAI-based text embedding models used to create vector encodings of each headline with principal component analysis (PCA) to exact the key features. The challenge of this work is to capture the time-dependent and time-independent, nuanced impacts of news on stock prices while handling potential lag effects and market noise. Financial and economic data were collected to improve model performance; such sources include the U.S. Dollar Index (DXY) and Treasury Interest Yields. Over 390 machine-learning inference models were trained. The preliminary results show that headline data embeddings greatly benefit stock price prediction by at least 40\% compared to training and optimizing a machine learning system without headline data embeddings. 
\end{abstract}

\begin{IEEEkeywords}
Machine Learning, Embeddings, Finance
\end{IEEEkeywords}

\section{Introduction}

Stock market prediction has long been a subject of interest for researchers, traders, and financial analysts. Traditional forecasting models primarily rely on historical price data and technical indicators to predict future movements. However, external factors influence financial markets, including macroeconomic indicators, investor sentiment, and breaking news. Among these, financial news and its headlines have proven to be a valuable source of information, as they reflect real-time market sentiment and can impact stock prices before traditional indicators respond.

This study focuses on leveraging headline data from the \textit{Wall Street Journal} to enhance stock price prediction for the SPDR S\&P 500 ETF Trust (SPY), an exchange-traded fund that tracks the performance of the S\&P 500, which consists of the largest 500 publicly traded companies in the United States. This work aims to improve the predictive accuracy of stock market movements on a daily timescale by incorporating news headline data.
We employ Principal Component Analysis (PCA) to extract meaningful features from news headlines [19]. These methods help identify key patterns in the data while reducing dimensionality and mitigating noise. Additionally, we integrate key economic indicators, such as the U.S. Dollar Index (DXY) and the United States Treasury PAR Yield Curve Rates, to further enhance the model's predictive performance [16], [17].
One of the main challenges in this study is addressing the time-dependent nature of news-driven stock market movements. News often has both immediate and delayed effects on stock prices, and market noise can obscure these relationships. Our approach seeks to balance the impact of news while accounting for potential lag effects in stock price reactions. 
In developing this work to examine the impact of news headline vector embeddings on stock price prediction, \textbf{390} machine learning models were trained.  
The rest of this paper is structured as follows: Section II provides background information on traditional stock market prediction techniques, the role of financial news, and the use of clustering and PCA in data analysis. Section III discusses the equipment used in this project. Section IV outlines the methodology, including data collection, feature extraction. Section V does into more detail regarding the model training and development methodology. Section VI presents the experimental results and analysis, while Section VII discusses future improvements and potential extensions of this work. Lastly, Section VIII discusses the conclusion with references. 

\section{Background}

Traditional methods rely on historical prices, but news sentiment plays a crucial role in market movements [1]. Studies show financial headlines improve prediction accuracy using embeddings and deep learning techniques [2]. PCA help reduce noise and extract key insights from news data for better modeling [3]. Capturing the time-dependent effects of news remains a key challenge due to market lag and volatility [1], [3].

\subsection{Brownian Motion and the Challenges in Stock Price Prediction}  
One of the key challenges in stock market prediction is accounting for the statistical Brownian motion inherent in stock price fluctuations [1], [9]. Stock prices are highly volatile and are often subject to random movements, making them difficult to predict using traditional machine-learning models. Optimizing a model to predict inherently random values would lead to a zero correlation between input and output data, rendering the model ineffective. This issue necessitates architectural decisions that mitigate the effects of randomness and enhance predictive reliability. One approach is introducing weighted loss functions and performance metrics that prioritize accurate predictions for significant price swings rather than minor fluctuations. The goal is not necessarily to perfectly predict absolute price changes but to accurately classify the market’s directional movement—whether the next day will yield an increase or decrease in stock price. By refining the model to focus on meaningful fluctuations rather than day-to-day noise, we can improve its practical applicability in financial forecasting.

\subsection{Headlines in the Machine Learning Pipeline}

One of the major goals of this project was to examine the impact of headlines on stock prices. As a result, many decisions were needed to quantify best and measure the impact of the headlines on the stock data. There are many different ways that short headlines could be integrated into the model, and commonly used techniques in literature include one-hot encoding, semantic analysis, and embeddings. 
There are advantages and disadvantages to using every technique, but embeddings were selected as just examining the text in headlines. With the rise of Generative Pre-Trained Transformers (GPT)-based Large Language Models (LLM), new techniques can extract meaning that can encode complex ideas in numbers. Headlines often contain or embody complex ideas that are abstracted in just a few words; examining the semantic feedback from these headlines may not result in a measurable impact. The main motive for using an embedding model was that clustering techniques and Principal Component Analysis (PCA) could be used to find structural relationships between headlines and their impacts [19]. This project brings one step closer to producing a mapping between stock prices and news headlines. 

However, there was no way to guarantee that the embedding model was not trained on the text of the headlines [20]. There was also no way to guarantee that the embedding model was trained on data exclusively between 1998 and 2015, the years defined as the training data for the model. As a result, embeddings may cause unwanted data leakage during model development.   

The choice of embedding model is essential in ensuring the accurate reduction of the conceptual ideas for use in our machine learning models. Two OpenAI models were tested: “text-embedding-3-small”  and “text-embedding-3-large” [20]. In testing, the difference resulted in the size and dimensionality of the output vectors, as well as the decimal precision of the embeddings. Another testing parameter was the reduced number of dimensions after the PCA process. A higher number of dimensions leads to easier data separation but runs into issues with data sparsity [19]. Data sparsity is the issue that occurs in vector point clouds, k-nearest neighbors search, and machine learning when there are not enough samples to saturate the input space sufficiently [19]. Problems such as poor generalization, overfitting, and difficulty estimating structures or distances emerge [14]. 

As a result of the headline embedding process, many categories with headlines were included. There were no situations where fewer headlines than days, but to handle more headlines, only one headline was chosen to stay as input to the model. 

\begin{figure}
    \centering
    \includegraphics[width=1\linewidth]{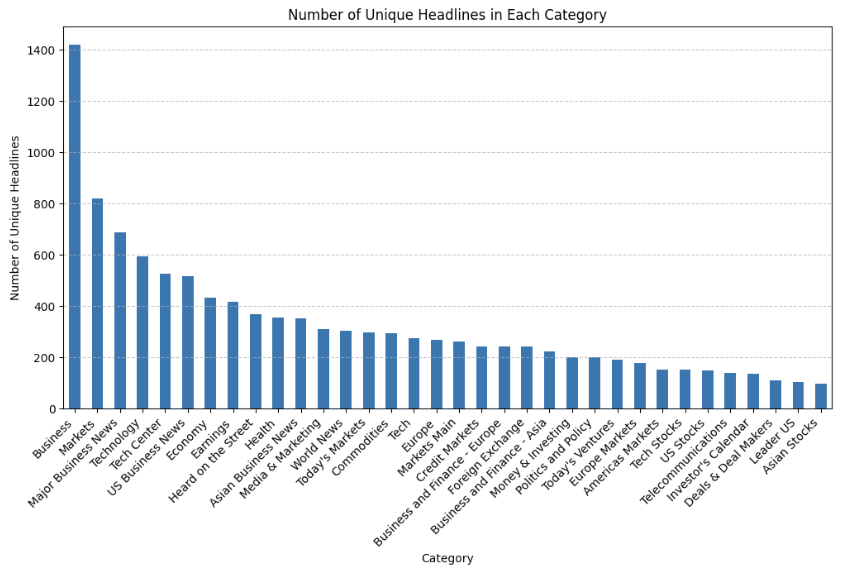}
    \caption{Number of Unique Headlines in WSJ Dataset}
    \label{fig:enter-label}
\end{figure}

\subsection{Explanation of Principal Component Analysis (PCA)}
This project used PCA to reduce the headline vector embeddings from the 1,536 dimensional original down to as low as two dimensions [19]. PCA identifies the principal components along which the data varies the most. The principal components in question are particular vector dimensions. This is achieved by computing the covariance matrix of the data and then determining the directions and magnitudes of variance [19]. The directions and magnitudes of variance are computed as the eigenvectors and eigenvalues [19]. The high-dimensional data is then projected onto a subspace of the largest eigenvectors, and only the most essential data is kept. The benefit of PCA is that it is computationally efficient, maintains the global structure of the data, and the variance captured is easily interpretable [19]. 

\subsection{Data Continuity}
Significant care and attention had to be taken to ensure that optimization and data sources in the model were accurate and reflected the data that can be collected today. For example, in the United States Treasury PAR Yield Curve Rates, many metrics did not exist continuously since 1998. These include the 1-month, 2-month, and 4-month yield curve rates, introduced on July 31, 2001, October 16, 2018, and October 19, 2022, respectively. 

There were also issues concerning the extremely high correlation of the input data sources. For example, in the U.S. Dollar Index (DXY) data, the DXY High and Low for each day were too highly correlated with the DXY Open, so the DXY High and Low were dropped from the dataset. The metric for high correlation was that the two time-series data cannot be more than 95\% correlated. This was calculated by computing the correlation matrix of the time series data and masking the upper triangle of the correlation matrix. As a result, the highly correlated columns emerged to be dropped. The goal of masking the upper triangle of the correlation matrix was that the redundant correlation values would not be double-counted, as the correlation matrix is symmetric. By masking the upper triangle, only the lower triangular values were examined to identify highly correlated features systematically. This step was essential to prevent multicollinearity, which can cause instability in the model’s weight updates and lead to misleading feature importance rankings. By eliminating features that were over 95\% correlated, the dataset was refined to include only the most informative and independent predictors, ensuring that the model could learn meaningful relationships rather than redundant patterns. 

Additionally, dropping highly correlated columns helped reduce the computational complexity of the model. With fewer redundant variables, the dimensionality of the dataset decreased, leading to faster training times and reducing the risk of overfitting. Removing these columns also improved the interpretability of the model, as each remaining feature contributed distinct information about stock market movements. Ultimately, this preprocessing step ensured that the model learned from diverse and complementary data sources rather than relying on excessive noise or redundant inputs.

Not all data sources had to be normalized like the stock data, and whether or not a time-series sequence should be normalized depended on the historical trends of that data. The decision on whether to normalize or not normalize depended on the distribution of points of a particular range, for example, between 1998 and 2015, which constitutes the training data. This is due to the way the DXY is designed. The DXY measures financial stability between the five of the world's financial currencies, including the US dollar. As it measures stability and is a relativistic metric, it is designed to remain within a particular range, unlike the absolute growth shown in the stock markets [21]. 
It is difficult for a model to try and predict the absolute value of a stock increase due to the limited number of data points around that range [20]. For example, a stock price of \$400 may seldom reach the same price again going forward. We can imagine a probability distribution where the target variable is the probability that a particular stock will increase or decrease from one day to another. This distribution of whether and how much a stock went up or down is skewed towards going up as it is not perfectly centered at zero. If it were perfectly centered at zero, there would be no need for normalization, but such modifications are essential for an exponentially increasing dataset over a long period.

\begin{figure}
    \centering
    \includegraphics[width=1\linewidth]{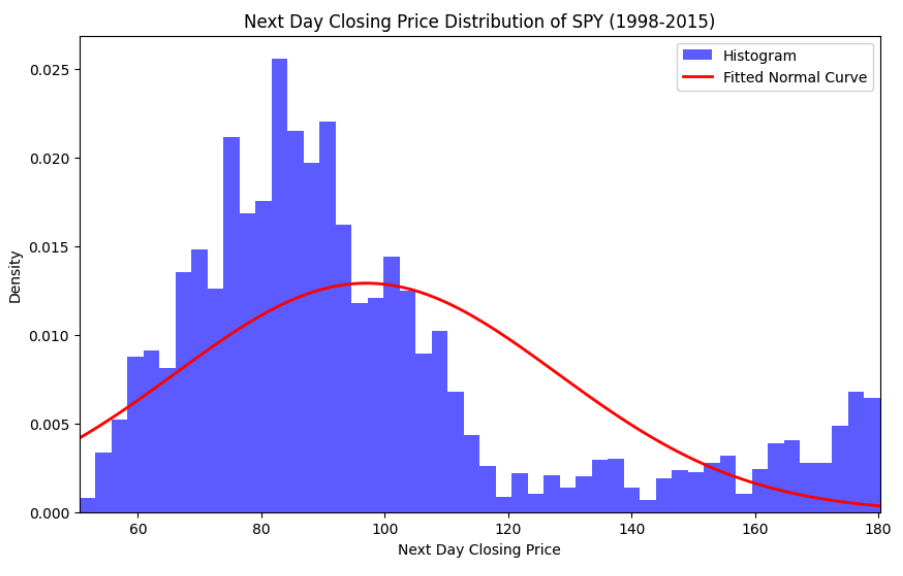}
    \caption{Distribution of SPY Close over time.}
    \label{fig:enter-label}
\end{figure}

\begin{figure}
    \centering
    \includegraphics[width=1\linewidth]{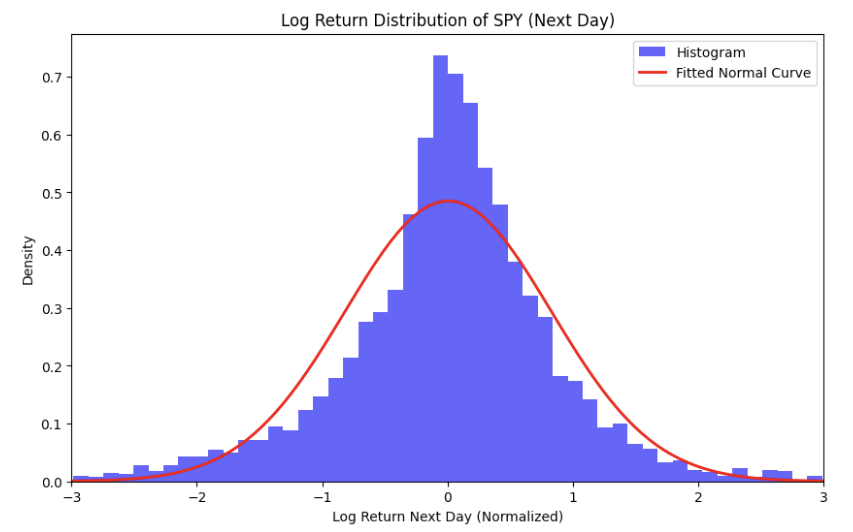}
    \caption{Normalized distribution of SPY Close.}
    \label{fig:enter-label}
\end{figure}

\subsection{Performance Metrics}
Although MSE is widely used, it has a significant drawback in heavily penalizing large deviations while giving disproportionate preference to perfect predictions [15]. This characteristic creates a problem in stock price forecasting, where achieving an exact match between predicted and actual values is nearly impossible due to inherent market randomness. A model that consistently predicts values close to the actual price but never perfectly matches them can still have a high MSE, making it seem less effective than it is. In practical applications, financial analysts may be more interested in minimizing average percentage error rather than squaring the error magnitudes. The strong preference of MSE for "perfect" predictions over "good enough" predictions can misrepresent the real-world usability of a model, requiring alternative metrics such as SMAPE.

To properly use all 20 years of data, much statistical analysis had to be performed to exhaust all possible possibilities for optimizing model performance. Many different performance metrics are used to optimize the performance and accuracy of the model, including  R², MSE and SMAPE [15]. Combining these three metrics in model development helped prevent the model from being overfitting and tailored for the job. 

Given the complexities of stock price prediction, a single performance metric cannot fully assess model performance. R² provides an overall measure of how much variance the model explains. MSE penalizes large errors to emphasize precision, and SMAPE offers a percentage-based error metric that is more intuitive for financial applications. These three metrics form a balanced evaluation framework, helping machine learning practitioners understand their models' accuracy and practical usability. The selection of an appropriate metric depends on the specific business use case: R² is ideal for explaining variance, MSE is effective when minimizing large deviations is critical, and SMAPE is best for ensuring practical interpretability of stock price predictions [15]. 

Machine learning models, particularly those used for financial forecasting and stock market predictions, require robust performance metrics to evaluate their accuracy, reliability, and generalization capability. Without proper evaluation metrics, it is impossible to determine whether a model is making meaningful predictions or simply overfitting historical data. Model evaluation becomes even more crucial in financial forecasting, where stock prices exhibit high volatility and often follow stochastic processes like Brownian motion. Metrics like R-squared (R²), Mean Squared Error (MSE), and Symmetric Mean Absolute Percentage Error (SMAPE) offer different perspectives on model performance, helping practitioners assess various aspects such as accuracy, consistency, and usability. Each metric serves a specific role: R² provides insight into how well the model captures variance, MSE penalizes large errors, and SMAPE accounts for percentage-based accuracy, making them complementary tools for understanding a machine learning model’s strengths and weaknesses.

The way all three metrics were calculated and used were as follows: 
 
\[
R^2 = 1 - \frac{SS_{res}}{SS_{tot}}
\]

where \(SS_{res}\) is the residual sum of squares (the total squared difference between actual and predicted values) and \(SS_{tot}\) is the total sum of squares (the total variance in the dataset). R² = 1 indicates a perfect model, while R² = 0 suggests that the model does not explain variance. In financial forecasting, R² helps assess whether the model captures meaningful patterns in stock price movements. However, R² does not directly penalize overfitting, meaning a model can achieve a high R² score while performing poorly on new, unseen data. Thus, it must be used alongside other metrics for a more holistic view of model performance.
  
Mean Squared Error (MSE) is one of regression-based machine learning models' most commonly used loss functions and performance metrics. It calculates the average of the squared differences between predicted and actual values:  
\[
MSE = \frac{1}{n} \sum_{i=1}^{n} (y_i - \hat{y}_i)^2
\]
where \( y_i \) represents actual values, and \( \hat{y}_i \) represents predicted values. The squaring of the errors ensures that larger errors have a disproportionately higher impact on the metric, making it particularly sensitive to extreme outliers. In stock market prediction, MSE helps quantify the magnitude of errors, allowing researchers to track how far off their predictions are on average. However, MSE alone does not provide an intuitive understanding of how well a model performs in practical financial settings because it penalizes larger deviations heavily, which might not always reflect real-world usability. Additionally, since MSE is measured in squared units, it does not provide an easily interpretable measure of prediction error relative to the actual stock price. This is where Symmetric Mean Absolute Percentage Error (SMAPE) becomes an essential metric. 
\[
\text{SMAPE} = \frac{100\%}{n} \sum_{i=1}^{n} \frac{|y_i - \hat{y}_i|}{\frac{|y_i| + |\hat{y}_i|}{2}}
\]
SMAPE expresses errors as a percentage of actual stock price changes, making it more interpretable and better suited for stock price predictions than raw MSE values. The MSE heavily penalizes extreme deviations, which can be misleading when dealing with volatile stocks that experience large fluctuations. In contrast, SMAPE scales errors relative to actual stock prices, ensuring that predictions are evaluated fairly regardless of the absolute price of the stock. For example, an error of \$5 on a \$10 stock should be considered worse than an error of \$50 on a \$500 stock, but MSE treats both errors as equivalent in absolute terms. SMAPE resolves this issue by normalizing errors relative to stock price movements, providing a more realistic and practically useful performance metric. 

\section{Equipment}
This project would not have been possible without extensive support from the open source software community. Table 1 discusses several software packages used, alongside the Python programming language version 3.10.12, to streamline the development of this machine learning model. All code was executed in Jupyter Notebooks and in Python modules on a Macbook M1 Pro laptop. 

\begin{table}[h]
    \centering
    \begin{tabular}{ll}
        \toprule
        \textbf{Software} & \textbf{Use In Project} \\
        \midrule
        NumPy & A library for numerical computing. \\
        Pandas & Structured data processing. \\
        Scikit-learn (sklearn) & Library for clustering. \\
        Yahoo Finance & Retrieve financial data from Yahoo Finance. \\
        Matplotlib & A visualization library. \\
        JSON & Build and read the Headlines dataset. \\
        Shutil & Duplicating datasets for recoverable operations. \\
        Threading & Optimize headline embedding, model training. \\
        Queue & Optimize headline embeddings. \\
        Multiprocessing & Optimize model training. \\
        PyTorch & An open-source deep learning framework. \\
        Optuna & Streamline hyperparameter optimization. \\
        Seaborn & Visualization for correlation matrices. \\
        \bottomrule
    \end{tabular}
    \caption{Software Descriptions}
    \label{tab:software_descriptions}
\end{table}

\section{Methodology}
Several economic and financial indicators were used to track stock performance. The SPY's daily opening and closing, volume, and volatility were used as financial indicators. This project had several goals; one was designing a machine learning system to predict the SPY's next-day performance. Another goal was to examine the statistical relationship and the benefit predictions made from including the headline data. 

Several economic and financial indicators were used as inputs to the model. A publicly available data set containing headlines from the Wall Street Journal was used as the composite for the headline data. This contained over 15,000 unique headlines spanning between 1998 and 2021. Several macroeconomic indicators were also used. These include the Dollar Stability Index (DXY) and the Federal Interest Yields Data. The goal was to predict the percent change in the stock price to account for daily fluctuations.

Several challenges were tackled regarding converting the Wall Street Journal headlines into an input the model can use. An OpenAI-based embedding model created an embedding vector point cloud with over 18 thousand headlines. This approach aimed to make the ideas and impacts encapsulated in the headlines searchable and relatable. Statistical methods could be used to build better categorization. For example, headlines that discuss significant economic regulation changes and significantly impact stock prices would not necessarily be encapsulated if a one-hot encoding technique was used to encode the headline data. Sentiment analysis was less preferable due to the advantage of using clustering and other tasks to build a correlation between the stock price and the headlines. The OpenAI embedding model consisted of a small and a large model that specialized in taking in a period of text and converting it into a vector that could be used in cases such as principal component analysis and clustering. The headlines met several parameters, such as the category of the specific news source and the date, which came in handy.

This model aims to predict the next day's stock price compared to the previous day. Two broad classes of models were developed: one with the Wall Street Journal headline embedding data and one without. Both models could be analyzed and compared to measure the specific impact of the headline data on real-world performance. 

Several machine learning architectures, design practices, and principles were used to develop both models. These practices include K-cross-fold validation and significant comparison analysis. Extensive model analysis was performed, testing several model architecture types, such as a feed-forward neural network and a Long Short Term Memory (LSTM) deep neural network architecture.

One challenge regarding the prediction of stock prices is that stock prices generally go upwards and do not come back down. For example, a stock price of \$100 in 1998 is acceptable, but the same value in 2021 is unacceptable and is likely an error. The fact that stocks tend upwards over long periods could introduce sources of error into the model and need to be accounted for. As a result, one technique was to train two more classes of models that experimented with two methods that only provided the relative change in the stock price rather than the absolute value of the stock price. With these new data sources, a model can predict and examine changes in a time-independent manner, recognizing structures within the data provided for that day. This leads to a model relying heavily on the embedding headline structure for that day.

Preventing data leakage was significantly important in this project. To be able to perform K-cross-fold validation while having a data set that could include data from other times was to split the data before training and testing. The total data collected was between 1998 and 2021. Training the model and performing K-cross-fold validation were performed on the data between 1998 and 2015. It was then tested extensively on data between 2016 and 2019, with the data from 2020 and 2021 serving as a reserve for further validation and testing.

\subsection{Model Architecture}
A significant innovation in this research was exploring how headline data can best enhance stock price prediction models. While machine learning techniques are crucial in this process, the primary focus was on integrating diverse data sources to create a model that provides real business value. A major challenge in financial forecasting is the inherent volatility of stock prices, which follow Brownian motion fluctuations daily. Thus, this research's key aspect was finding ways to utilize weakly correlated or seemingly uncorrelated data sources and extracting meaningful predictive insights. Machine learning models excel when trained on large, consistent datasets, but financial data is often limited, making it difficult to build robust models. This research prioritized optimizing data usage, applying innovative preprocessing techniques, and leveraging structured data sources creatively to improve predictive performance. Deep learning methods were deliberately minimized to reduce the risk of overfitting, as even a simple neural network architecture demonstrated a dropout rate of nearly 30\%, indicating a high degree of correlation among data points and the need for regularization to prevent overfitting.

\begin{figure}
    \centering
    \includegraphics[width=.75\linewidth]{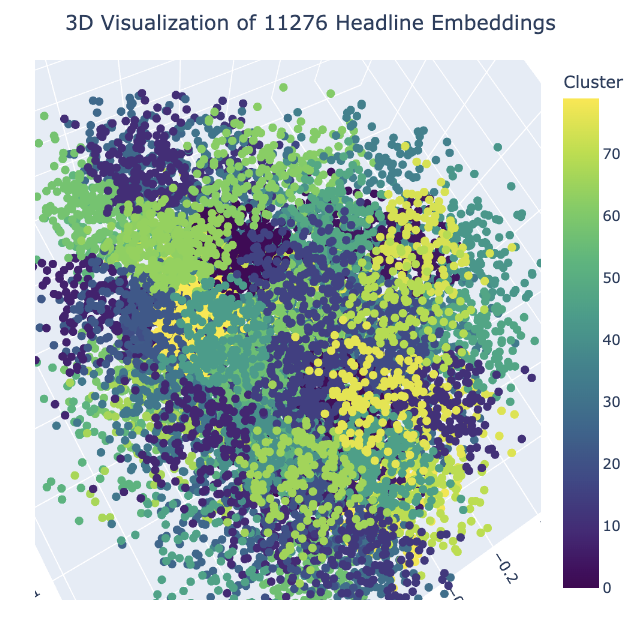}
    \caption{Visualization of the headline embeddings as a vector point cloud.}
    \label{fig:enter-label}
\end{figure}

Many different types of model architecture were used, including polynomial regression, a feed-forward neural network, and a deep neural network with a long- and short-term memory module. These three were chosen as primary to balance model complexity and overfitting. Too large and complex models tend to be prone to overfitting, whereas too simple models can sometimes lead to underfitting. Overfitting can occur when a model memorizes the output data, leading to low generalizability [8]. Underfitting can lead to poor performance during training and testing of the machine learning system [9]. 

Many optimizations were included to reduce data leakage, defined as future points that inadvertently affect past points. For example, there can be a situation where the next day's stock price can inadvertently be present as an input to the model when it should only ever be considered an output. This could lead to a situation where the model has information it is not supposed to have. When used in a real-life situation to predict the most recent stock price, the biggest problem emerges due to a lack of information. The model cannot just necessarily memorize a sequential time-variant sequence. One architecture design should be able to handle all matters of optimizations and data points regarding the prediction of numbers on a time series [9]. 

One of the main goals was to try and decouple the time series data to read that of a not time series data. One key item behind this is to randomize the order of the data. For the LSTM model, this randomization had to be turned off as the model is designed to make fine adjustments. Data from moving averages, such as 30-day or 7-day averages, were not generated and included as inputs to the model to reduce prior dependencies. One significant dependency was the inclusion of the log return calculation. 

It takes the stock price on the previous day and the next day to calculate the log return. As this ties two data points in time together, significant care was taken to prevent this mechanism from causing data leakage.

The goal of implementing a log-return calculation was to ensure that the model is optimized for predicting the best time to buy and sell stocks. Such forecasting is a secondary research goal for many model architectures and data processing types. 

\[
\text{Log Return}_t = \ln\left(\frac{P_t}{P_{t-1}}\right)
\]

\[
\text{Absolute Difference Return}_t = P_t - P_{t-1}
\]

\begin{figure}
    \centering
    \includegraphics[width=1\linewidth]{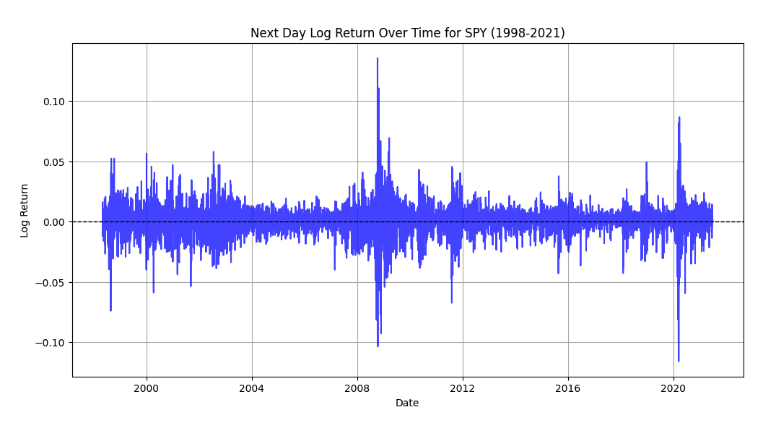}
    \caption{Log Return of SPY over time.}
    \label{fig:enter-label}
\end{figure}

After computing the Log Return of SPY, we see that the Signal-to-Noise Ratio (SNR) is extremely small at 0.024. This compares to the absolute pre-normalized signal with an SNR of 1.733. Interestingly enough, the Log Return signal of just noise is what we intend to have the model predict, as this is the precise noise that, over days and months, defines the entire distribution of the SPY stock signal. There are advantages to designing a model that can predict noise. In this case, it's important to note that the input variables to the model, at least those statistically correlated with the output, such as the current day's stock price, are just as noisy as the output variable [2]. 

Figure 3 describes the problem of pairs of vectors where one vector is the previous day, and the other is the next day. The challenge is to build a machine learning system that can separate the two points classes. However, the even distribution of input and output variable pairs of vectors makes it difficult to make generalizations as the distribution is not cleanly linearly separable. It is also important to note that the Shannon Entropy remains roughly the same before and after the Log Return computation, approximately 12.50 for both. 

\begin{figure}
    \centering
    \includegraphics[width=0.75\linewidth]{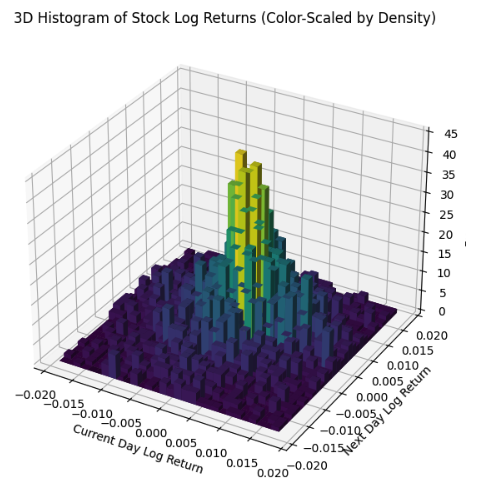}
    \caption{3D Histogram of Stock Log Returns}
    \label{fig:enter-label}
\end{figure}

\subsection{Comparing Time-Dependent and Time-Independent Models}
The research evaluated four broad classes of machine learning models:

\begin{itemize}
    \item Time-independent model without headline data
    \item Time-independent model with headline data
    \item Time-dependent model without headline data
    \item Time-dependent model with headline data
\end{itemize}

In the project, five major neural networks were tested. These architecture types included Gated Recurrent Units (GRU), Hidden Markov Model (HMM), Long-Short Term Memory (LSTM), Temporal Convolutional Networks (TCN), and a Feed-Forward Neural Network (FFNN). Each neural network was tested with three versions of the training data: time-dependent, logarithmic difference time-independent, and linear difference time-independent. Each version of the training data had two headline embedding variations, with 12 and 14 different Principal Component Analysis (PCA) dimensionality reductions tested. In total, there were \textbf{390} machine learning models trained. 

Every model was trained with 500 epochs. Training data was set between 1998 and 2015, and testing data was set between 2016 and 2019. 
Data processing techniques led to three classes of data, each used to create a new model. The time-dependent data was processed as a standard time-series system. There were two time-independent data sources created. These are the time-independent logarithmic return computation and the time-independent difference. This was where the linear difference between the previous day and today's stock was computed instead of the logarithmic return.

Principal Component Analysis (PCA) was used to reduce the dimensionality of the headline embeddings from the original 1,536 dimensions down to just two. Dimensionality reduction was tested on the logarithmic two, leading to eleven variations of headlines. If the data type where no headlines are included, there are twelve total data variations per class of data after the processing technique. 

Only one stock was tested, the SPY, as it is a composite of many different stocks that follows the composition of the S\&P 500. The news headlines were not biased or favoring any one particular company between 1998 and 2021. 

By training 390 machine learning inference models, the impact of news headlines on the next day's stock price could be quantified. With three data types, the impact of the headline on the model's ability to make a decision became clear. As hyperparameter optimization was used to extract the best performance out of each of the 390 machine-learning models, the true impact of headlines on the next day's stock price can be examined. Further improvements could be testing different stocks, as well as testing different headline sources. 

\subsection{Discussion of Model Architectures}
In the project, five major neural networks were tested. These architecture types included Gated Recurrent Units (GRU), Hidden Markov Model (HMM), Long-Short Term Memory (LSTM), Temporal Convolutional Networks (TCN), and a Feed-Forward Neural Network (FFNN). All were implemented using PyTorch. This subsection discusses each model architecture. All model weights were initialized using Xavier uniform initialization. The purpose of this was to ensure stable gradient flow during training. All models included dropout layers to help prevent overfitting. Most models included a flexible linear layer that mapped the varying input dimensionality into a lower dimensional layer for standardized learning.  

The Feed-Forward Neural Network (FFNN) implementation consisted of a linear input layer feeding into a series of hidden layers. The number of hidden layers was a hyperparameter tuned for each model, but the best performing in the research was with the number of hidden layers set to one with its size set to 32 cells.   

Improving upon the FFNN implementation was the Long-Short Term Memory (LSTM) implementation. Similar to the FFNN, the first linear layer maps the input dimension into the hidden layer dimension. 

The Gated Recurrent Units (GRU)-based neural network was selected due to the GRU's proven efficiency in capturing time-dependent changes and advantages over traditional Recurrent Neural Networks (RNN). The model consists of a series of layers, starting with a linear layer with ReLU activation and then into the GRU layer. The GRU layer selects the last timestamp to predict the upcoming timestamp. A final linear layer reduces the GRU output to the target prediction horizon. 

The Temporal Convolutional Network (TCN) implementation introduces a convolutional approach to sequence modeling, utilizing dilated causal convolutions to capture temporal dependencies without relying on recurrent structures. The model is built from a stack of TemporalBlock modules, each composed of two 1D convolutional layers with increasing dilation rates, interleaved with ReLU activations and dropout layers. After each convolution, causality is enforced by removing excess padding from the output. Residual connections are applied across each temporal block to promote gradient flow and improve training stability. As with the other models, the input first passes through a linear layer for dimensional alignment, and the output from the final temporal block is reduced to the target prediction size through a linear layer acting on the final time step's output. The dilation pattern used in the convolutional stack enables the model to efficiently model long-range dependencies in the input sequence. 

The Neural Network Hidden Markov Model (NN-HMM) was implemented to simulate a simplified version of a Hidden Markov Model, with the state transitions abstracted and the emission probabilities learned via feed-forward neural networks. The model defines a fixed number of hidden states, each associated with its emission network—a multilayer perceptron mapping the input features to the prediction target. The input is processed only at the last time step, aligning with the behavior of other models like the LSTM and GRU. Each emission network produces an independent prediction and a weighted average at inference time. The weights are learned log probabilities, transformed via a softmax to ensure they sum to one and are consistent across the batch. This approach enables the model to simulate latent state behavior while leveraging the representational capacity of neural networks, offering a flexible alternative to classical HMMs. Implementing a hybrid model rather than a full HMM provided flexibility when using the preexisting toolings to judge all other model architectures evenly on the same level. 

\subsection{Challenges of Hyperparameter Optimization}
The hyperparameter tuning process in this research was highly effective in optimizing the model's performance in various scenarios. PyTorch was also used to fine-tune hidden layer sizes, dropout, learning, and batch sizes, ensuring all models were configured for maximum accuracy and stability. However, one remaining challenge is the issue of data reusability: Although the model performs well on general stock predictions, its applicability to specific stocks remains an open question [4]. One possible improvement is transfer learning—where a pre-trained model on general stock data is further fine-tuned for individual stocks like Microsoft or Apple. This technique has been highly effective in computer vision and natural language processing (NLP) and could be adapted to financial modeling [8]. Such an approach would allow the model to retain broad market insights while being customized for specific stocks, improving prediction accuracy for targeted investments.

\section{Model Development Methodology}
Training 390 machine learning inference models in an organized manner required strictly following the streamlined machine learning development pipeline. 

\begin{figure}
    \centering
    \includegraphics[width=1\linewidth]{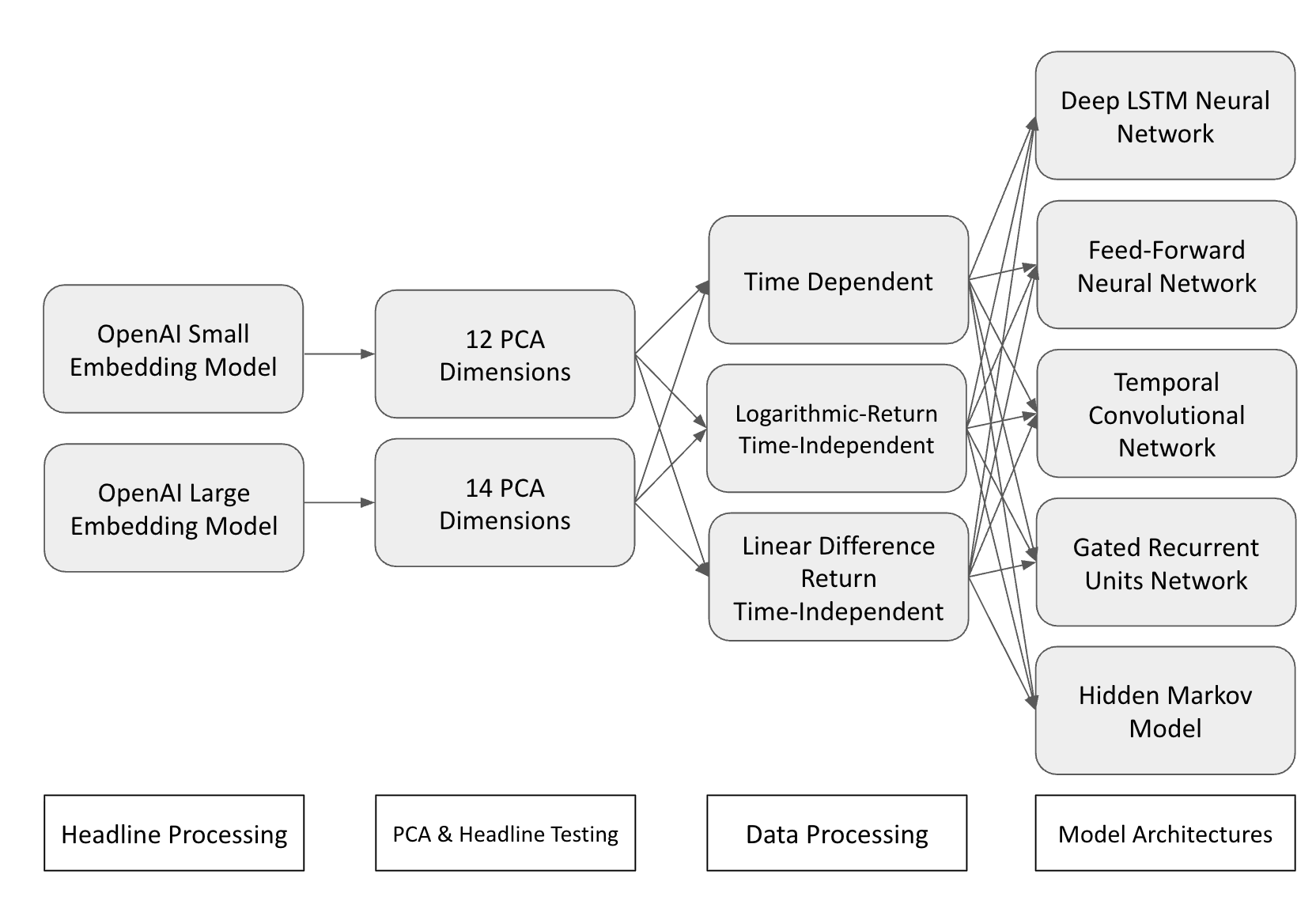}
    \caption{Diagram of all combinations of models tested.}
    \label{fig:enter-label}
\end{figure}
The data-to-model pipeline contained five steps following the machine learning development lifecycle. They are: 
\begin{itemize}
    \item Exploratory Data Analysis
    \item Data Pre-Processing
    \item Dataset Feature Extraction
    \item Model Development, Training and Testing
\end{itemize}

\subsection{Exploratory Data Analysis and Pre-Processing}

In the first step, all datasets are loaded and examined with various open-source tools. The library matplotlib was used to plot the data sources as a function of time. 

\begin{figure}
    \centering
    \includegraphics[width=1\linewidth]{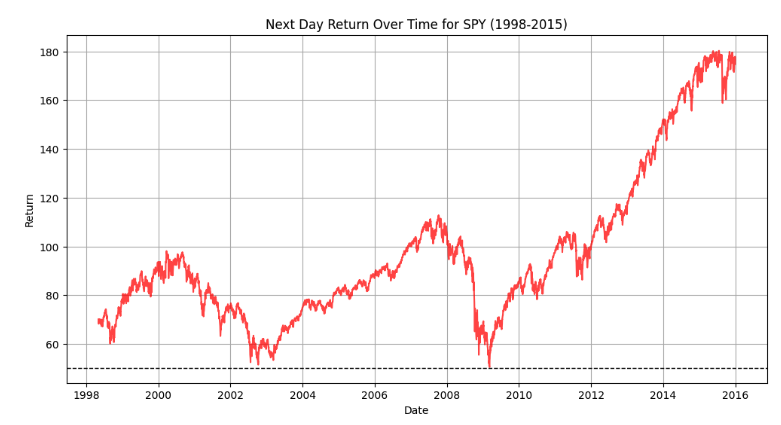}
    \caption{Next Day Return for SPY as a function of time.}
    \label{fig:enter-label}
\end{figure}

\begin{figure}
    \centering
    \includegraphics[width=1\linewidth]{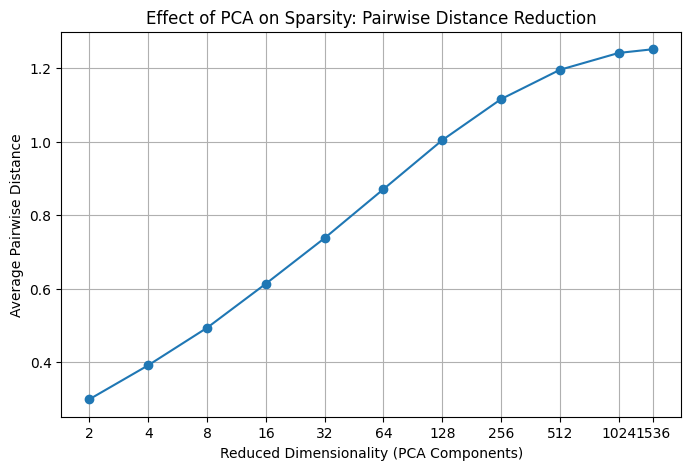}
    \caption{Pairwise Distance Reduction as a result of PCA.}
    \label{fig:enter-label}
\end{figure}

In this step, dropping any data points that do not exist across all input datasets was vital. This is important as there are situations where, for example, the markets are closed on one day, yet there is data from another data source, such as yield data. The resulting decision was that no prediction can be made on days without data for the SPY, which translates to days the New York Stock Exchange (NYSE) is closed. One of the key features is the previous stock price itself, so such changes are necessary. 
Another assumption was that any gaps in the market where the market is closed are assumed not to have an impact. This was necessary to handle the sheer rollover from Friday's price to Monday. 

\subsection{Dataset Feature Extraction}

\subsubsection{Yields Data}
The Federal Yields Interest Rate data were included as inputs to the model. In particular, the 3-month and 10-year Interest Rate data were selected out of the number of yield rates available. Such a decision maintained several advantages. Firstly, not all of the Yields Interest Rate data existed continuously since 1998, as some Yields data was introduced later on, such as the 4-month yield rate. For those that did exist continuously since 1998, as there was over 95\% correlation between most of the Federal Interest Rate Yields data, the decision was only to include two data sources. Including highly correlated input data would later lead to matrix invertibility issues when performing backpropagation. 

\subsubsection{Dollar Index Data}
Similar to the Federal Yields Interest Rate Data, the DXY Low and High for each day had over 95\% correlation with each other and had to be dropped as inputs to the model [16]. Instead, only the DXY Open was provided as input for the model. 

\subsubsection{Headlines}
This step focused on creating the headline embeddings and processing the time-series data for use in the ensemble of machine learning models. Properly filtering the headlines was necessary to ensure the input values were relevant to the use case. Although over eighteen-thousand headlines were provided from the Wall Street Journal headline dataset, only a subset directly discuss financial markets or a particular stock performance, a little over eleven-thousand [18]. This results in a database size of over 700 megabytes without compression. 

Not all days had a headline that were related to finance, business, or other relevant categories, and most days had two or three headlines for each. With too many headlines, one was chosen randomly to be placed into the model. Such a process had advantages in reducing biases, as there was no guarantee that a particular headline would have the strongest impact on stock prices out of the ones placed on that day. 

As there were multiple headlines for a particular day, a decision had to be made on which headline to include in the model. As these headlines in embedding form encode ideas, it could be fair to add all vectors together and then normalize to create a vector containing the "average" of the ideas in these headlines. However, this would introduce artificial data into the dataset, as on most days, only one embedding was directly created from a real financial headline. Hypothetically, if an inverse function could take the full embedding and convert it back into a textual output mapping, no data loss would occur if the original headline is not subtracted or added in any form. Suppose headlines that now encode ideas are added and subtracted together. There is no guarantee that in the very high dimensional space the sum of the two embeddings still reproduces a relevant embedding in the text space in the inverse conversion [12]. Such a problem also appears during dimensionality reduction to reduce data sparsity. 

\subsubsection{Headline Dimensionality Reduction}
Principal Component Analysis (PCA) was used to compress the headlines from the 1535 dimensions down to dimensions that can be used for the Machine Learning model. The concept behind such an operation is that two ideas in space in a 1,535-dimensional space are extremely unlikely to be "near" each other, making mappings in a machine-learning model more prone to overfitting. This analysis aimed to reduce the high dimensional embedding data into lower dimensional data to reduce sparsity. Much experimentation was done to test the best configuration to maximize model accuracy and performance. The PCA step reduced the 1,535-dimension vector to a set of reduced dimensions to reduce issues related to data sparsity. Different dimensionality reductions were tested: 2, 4, 8, 16, 32, 64, 128, 256, 512, 1024, and the original dimensionality of 1535 to test the model's performance without PCA. 

A new and unique model was trained for every combination of PCA dimensionality reduction. The dimensionality of the data directly from the OpenAI model output was a 1,536 dimension vector per embedding with at most 20 decimals of precision, which in the research proved to strike a balance between precision and computation speed. 

The results showed that including relevant financial headlines from the Wall Street Journal significantly improved model prediction performance by up to 40\%. Extensive comparative analysis was performed. 

Preliminary results indicate that incorporating headline data embeddings significantly improves predictive performance by at least 40\% compared to models relying solely on historical stock prices and economic indicators. These findings suggest that financial news plays a crucial role in market forecasting and can be effectively utilized in stock prediction models.

The results indicated that time-independent models provided greater flexibility in handling diverse market conditions. In contrast, time-dependent models showed better sequential trend recognition but suffered from data sparsity and reduced generalization. Given the randomness of stock market fluctuations, the risk of overfitting in a time-dependent model was high [6]. Ideally, we want a model that learns inherent patterns in the market rather than memorizing past data. However, the time-independent approach comes with challenges, since the model never sees the same data point twice, it becomes harder to extract deeper insights from limited datasets. Future research could explore hybrid approaches that leverage the advantages of both time-dependent and time-independent models to improve accuracy without overfitting.

\section{Results}
\begin{figure}
    \centering
    \includegraphics[width=1\linewidth]{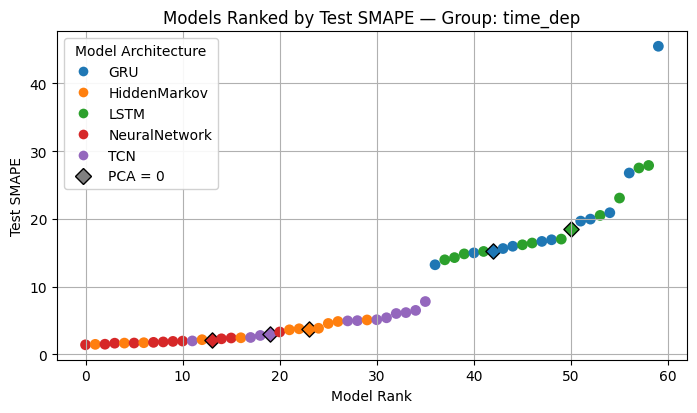}
    \caption{Ranking of model performance by SMAPE}
    \label{fig:enter-label}
\end{figure}

The results of the time-independent operations proved to be just as good, if not better, as time-dependent ones. By forcing the model to examine the data source shuffled right before it enters the model, it cannot make the same generalization that a feed-forward no network would be able to make. Without the proper optimizations, a model that would be placed in the situation would be forced to create worse generalizations. Now, as there are over 4000 unique days in the data set, it is essentially saying that if the model can endure random variations in the time sequence, This model can handle every single unique combination of these 4000 days, which leads to an ability to create more use out of a certain data set. Not all three models have the same architecture. The only difference was the date and time independence. So optimizations could have been made trying to approve the specific benefit of being able to approve a particular day. Still, one of the benefits of such a resilient and effective no-network architecture is that Such authorizations do not necessarily have to be pursued to the same extent. Many different classes of models or tested with a vast array of principal component analysis dimensions. So many combinations of models were created to see the best operation to perform on the data to get the best performance out of this specific known architecture. And class of architecture. They were number combinations based on the number of dimensions from the principal analysis, as well as the type of data transform prefer performed on the time series data to either make it time in variant or not, which led to a very large number of models that had to be tested and trained, per Machine Learning model architecture.

\begin{figure}
    \centering
    \includegraphics[width=1\linewidth]{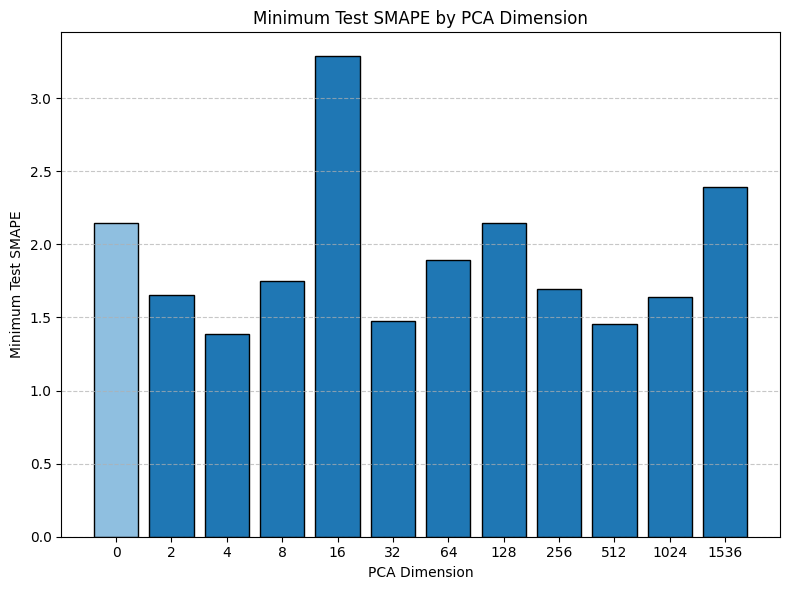}
    \caption{Minimum SMAPE achieved for each PCA dimension (across all models)}
    \label{fig:enter-label}
\end{figure}

\begin{table*}[t]
\centering
\caption{Best Models Sorted by SMAPE (Small OpenAI Embedding Model)}
\begin{tabular}{lrrrrrrr}
\toprule
Architecture & PCA Dimension & Train MSE & Test MSE & Train SMAPE & Test SMAPE & Train  R²& Test  R²\\
\midrule
NeuralNetwork & 4 & 0.000302 & 0.001212 & 5.716504 & 1.390008 & 0.993723 & 0.993262 \\
HiddenMarkov & 512 & 0.000324 & 0.001229 & 5.775349 & 1.455149 & 0.993260 & 0.993032 \\
NeuralNetwork & 32 & 0.000284 & 0.001252 & 5.594068 & 1.474592 & 0.994116 & 0.993163 \\
NeuralNetwork & 512 & 0.000674 & 0.001384 & 6.082242 & 1.626646 & 0.986104 & 0.992324 \\
HiddenMarkov & 1024 & 0.000337 & 0.001376 & 6.220379 & 1.640225 & 0.992995 & 0.992307 \\
NeuralNetwork & 2 & 0.000287 & 0.001316 & 5.470101 & 1.651083 & 0.994050 & 0.992689 \\
HiddenMarkov & 256 & 0.000199 & 0.001385 & 4.489222 & 1.697381 & 0.995867 & 0.992307 \\
NeuralNetwork & 8 & 0.000287 & 0.001466 & 5.423363 & 1.750581 & 0.994057 & 0.992123 \\
NeuralNetwork & 256 & 0.000580 & 0.001468 & 5.898670 & 1.811211 & 0.987938 & 0.991862 \\
NeuralNetwork & 64 & 0.000337 & 0.001702 & 5.763248 & 1.889726 & 0.993005 & 0.991113 \\
\bottomrule
\end{tabular}
\end{table*}

The second model is a time-dependent model designed to predict one-day-ahead on data with its proper temporal sequence. This model is designed to capture short-term dependencies by using contextual information from prior headlines. It can learn temporal causality, aligning more closely with real-world market behavior where short-term memory and news momentum often influence prices [4], [6]. However, this added temporal complexity makes the model more susceptible to overfitting transient or spurious patterns, especially in volatile environments [8]. Additionally, the training process becomes more intricate, requiring careful feature engineering and time-aware validation strategies [9].

The third model extends the time-dependent approach to a longer-term forecasting horizon, predicting stock movements one week ahead using a larger temporal context of headline embeddings. This model is motivated by the idea that the market interpretation of information can evolve gradually, and certain headlines have delayed or compounded effects. It is particularly suited for modeling long-term sentiment drift and thematic influence [12]. Such models may also align better with intermediate-term investment strategies that seek to avoid short-term noise [3]. However, the trade-off for this temporal breadth is a lower signal-to-noise ratio; the further into the future, the prediction, the more diluted the explanatory power of any single headline becomes [10]. This necessitates using larger and more carefully curated datasets to maintain predictive accuracy.

The three modeling strategies reflect different assumptions about how news influences markets. The time-invariant model identifies persistent semantic patterns but lacks awareness of dynamic context. The short-term time-dependent model incorporates sequential information, making it well-suited to capturing immediate reactions to news while also prone to overfitting if not carefully managed. The long-term model accommodates delayed and cumulative effects, offering strategic value for broader forecasting horizons but at the cost of reduced granularity and increased data demands. The choice of model should ultimately be driven by the prediction objective, the expected temporal profile of news impact, and the tolerance for volatility inherent in the target application. 

\section{Further Work}
One of the key implications of this research lies in its ability to transform a time-dependent sequence of data into a time-independent representation. This enables the model to make time-agnostic predictions while capturing temporal patterns, thereby improving generalization and repeatability through higher-order pattern recognition \cite{b6, b9, b15}. Future development includes deploying this model onto custom hardware modules to accelerate inference and computation speeds dramatically. Since PyTorch models can be exported as compilable C code, this provides a pathway toward implementing the model at a hardware level, including potential exploration of SystemVerilog-based architectures for ultra-fast embedded machine learning applications.

A central motivation of this work is minimizing the latency between data acquisition and actionable model output. In fast-moving environments like financial trading, reducing this latency is critical. With sufficient financial investment and infrastructure, this system could enable rapid decision-making for applications such as micro-trading \cite{b1, b3, b11}. Moreover, the inverse of this modeling approach could have value for organizations such as The Wall Street Journal, allowing them to quantify how much their headline structure influences market behavior, providing insight into the complex relationship between media sentiment and stock price movements \cite{b10, b13, b14}.

Although causation and correlation remain difficult to fully unravel \cite{b4, b8}, the broader implication of this work lies in its ability to draw meaningful connections between news data and stock price changes, which is a central challenge in modern statistical and machine learning applications \cite{b5, b12, b15}.

There is also substantial potential to enhance model performance by expanding the data sources and timeframes. This study was limited to financial headlines between 1998 and 2021, primarily due to access restrictions and API limitations from providers like The Wall Street Journal \cite{b18}. Future iterations could focus on integrating multiple financial news sources, enabling more granular and diverse insights across a wider set of companies and industries. For example, collecting company-specific news — such as Microsoft’s press coverage — could allow for more targeted prediction models \cite{b1, b3, b14}.

Building a robust data pipeline to collect, merge, and clean multiple headline sources would be a valuable next step. Doing so would help address limitations in data coverage and improve model accuracy in stock prediction tasks.

Finally, developing machine learning models optimized for embedded systems opens an exciting frontier. These systems offer the potential for fast, efficient, and accurate predictions directly on specialized hardware, a promising direction for the broader field of machine learning-driven stock prediction.

\section{Conclusion}
The challenge of predicting the next day's stock price can be improved by including relevant financial headlines. The methods by which relevant financial headlines can be included in a machine learning model consist of using an embedding model to build a vector cloud and then reducing the dimensionality to reduce data sparsity. This paper documented five model architectures and three different data processing techniques to show that including financial headline data can improve stock price prediction performance by reducing errors from 35-40\%. By designing particular machine learning models with a precise methodology and experimental framework for rapid experimentation, inference models can be optimized and improved in predicting stocks.

\newpage
\section{Appendix}
\subsection{Optimizing Headline Calls}
With over 18,000 calls to the OpenAI API for creating headline embeddings, an efficient process is needed to reduce the time taken from hours to minutes. The key innovation in Python was to parallelize the calls by using Jupyter Notebook and Threading libraries. This made it possible to have many virtual workers do the work in parallel rather than one sequential worker. 

To process a headline takes three steps: First, the headline in question must be prepared to send to OpenAI. Second, a worker must send and wait for the result of this call. Lastly, the resulting embedding must be saved to disk in the database. Due to the short length of the headline text data, as well as the efficiency of the OpenAI embedding model, each call to OpenAI took approximately 0.04 seconds to receive the embedding for that headline. Hundreds of worker threads in parallel could make calls to OpenAI and store the embedding result in memory temporarily. The bottleneck comes with writing to the database and preventing read-and-write conflicts. The solution was to create a cache, where writes to the database were only performed once a certain threshold was met from the embedding result. For example, the code would have 500 worker threads all making calls to the OpenAI API, but there were only four writing threads. Once every 100 results from the OpenAI API did a write thread execute and write 100 results to the database at once. 

Further work in this section can use a database that is more resilient than a JSON structure, as well as work on local duplication to prevent threads from locking up due to there only being one final database to which all calls were written. 


\begin{thebibliography}{00}

\bibitem{b1} 
Stock Movement Prediction with Financial News using Contextualized Embedding from BERT, arXiv, 2021.

\bibitem{b2} 
Deep Learning for Stock Market Prediction Using Event Embedding and Technical Indicators, IEEE Xplore, 2018.

\bibitem{b3} 
A Deep Fusion Model for Stock Market Prediction with News Headlines, Springer, 2024.

\bibitem{b4} 
D. H. Bailey, J. M. Borwein, M. López de Prado, and Q. J. Zhu, "The probability of backtest overfitting," \textit{Journal of Computational Finance}, vol. 20, no. 4, pp. 39–69, 2016.

\bibitem{b5}
L. Breiman, "Random forests," \textit{Machine Learning}, vol. 45, no. 1, pp. 5–32, 2001.

\bibitem{b6}
T. Fischer and C. Krauss, "Deep learning with long short-term memory networks for financial market predictions," \textit{European Journal of Operational Research}, vol. 270, no. 2, pp. 654–669, 2018.

\bibitem{b7}
M. Grootendorst, "BERTopic: Neural topic modeling with class-based TF-IDF," \textit{arXiv preprint arXiv:2203.05794}, 2022.

\bibitem{b8}
C. R. Harvey and Y. Liu, "Backtesting," \textit{Journal of Portfolio Management}, vol. 42, no. 1, pp. 13–28, 2015.

\bibitem{b9}
R. J. Hyndman and G. Athanasopoulos, \textit{Forecasting: Principles and Practice}, OTexts, 2018.

\bibitem{b10}
B. Kelly, S. Pruitt, and Y. Su, "Historical textual analysis and stock returns," \textit{Journal of Financial Economics}, vol. 142, no. 3, pp. 1290–1310, 2021.

\bibitem{b11}
C. Krauss, X. A. Do, and N. Huck, "Deep neural networks, gradient-boosted trees, random forests: Statistical arbitrage on the S\&P 500," \textit{European Journal of Operational Research}, vol. 259, no. 2, pp. 689–702, 2017.

\bibitem{b12}
A. K. Nassirtoussi, S. Aghabozorgi, T. Y. Wah, and D. C. L. Ngo, "Text mining for market prediction: A systematic review," \textit{Expert Systems with Applications}, vol. 41, no. 16, pp. 7653–7670, 2015.

\bibitem{b13}
P. C. Tetlock, "Giving content to investor sentiment: The role of media in the stock market," \textit{Journal of Finance}, vol. 62, no. 3, pp. 1139–1168, 2007.

\bibitem{b14}
Y. Zhang, S. Skiena, et al., "Trading strategies with news sentiment: A revisit with transformer-based language models," in \textit{Proc. 1st Workshop on NLP and Financial Technology (FinNLP)}, 2020.

\bibitem{b15}
G. James, D. Witten, T. Hastie, R. Tibshirani, and J. Taylor, An introduction to statistical learning: With applications in Python. Springer Cham, 2023. [Online]. Available: \url{https://www.statlearning.com}.

\bibitem{b16}
Balabaskar, "\textit{US Dollar Index Data (2001–2022)}," Kaggle. [Online]. Available: \url{https://www.kaggle.com/datasets/balabaskar/us-dollar-index-data/data}

\bibitem{b17}
U.S. Department of the Treasury, "\textit{Interest Rate Data - CSV Archive}," [Online]. Available: \url{https://home.treasury.gov/interest-rates-data-csv-archive}

\bibitem{b18}
A. Joshi, "\textit{WSJ Headline Classification Dataset}," Kaggle. [Online]. Available: \url{https://www.kaggle.com/datasets/amogh7joshi/wsj-headline-classification}

\bibitem{b19}
"Mathematics for Machine Learning". Copyright 2020 by Marc Peter Deisenroth, A. Aldo Faisal, and Cheng Soon Ong. Published by Cambridge University Press. [Online] Available: \url{https://mml-book.github.io/}

\bibitem{b20}
OpenAI Embeddings \url{https://platform.openai.com/docs/guides/embeddings}

\bibitem{b21}
"U.S. Dollar Index (USDX)," Investopedia. [Online]. Available: \url{https://www.investopedia.com/terms/u/usdx.asp}


\end{thebibliography}
\end{document}